\documentclass[prb,twocolumn,superscriptaddress]{revtex4}

\usepackage{graphicx}
\usepackage{pstricks}
\usepackage{pst-node}
\usepackage{amsmath}
\usepackage{bbold}
\usepackage{epsfig}
\usepackage{algorithm}
\usepackage{algorithmic}

\preprint{LA-UR-08-05038}

\begin{document}

\title{Cache oblivious storage and access heuristics for blocked matrix--matrix
multiplication}

\date{\today}

\newcommand{\kth}{Department of Theoretical Chemistry, School of Biotechnology,
Royal Institute of Technology, SE-10691 Stockholm, Sweden}
\newcommand{\lanl}{Theoretical Division, Los Alamos National Laboratory, Los
Alamos, NM 87545, USA}

\author{Nicolas Bock}
\email[]{nbock@lanl.gov}
\affiliation{\lanl}

\author{Emanuel H. Rubensson}
\email[]{emanuel@theochem.kth.se}
\affiliation{\kth}

\author{Pawe{\l} Sa{\l}ek}
\affiliation{\kth}

\author{Anders M. N. Niklasson}
\affiliation{\lanl}

\author{Matt Challacombe}
\affiliation{\lanl}

\begin{abstract}

We investigate effects of ordering in blocked matrix--matrix multiplication.  We
find that submatrices do not have to be stored contiguously in memory to achieve
near optimal performance. Instead it is the choice of execution order of the
submatrix multiplications that leads to a speedup of up to four times for small
block sizes.  This is in contrast to results for single matrix elements showing
that contiguous memory allocation quickly becomes irrelevant as the blocksize
increases.

\end{abstract}

\maketitle

\section{Introduction}

In current state--of--the--art algorithms for large scale electronic structure
calculations probably the most important operation is sparse matrix--matrix
multiplication. To name but a few important applications, sparse matrix--matrix
multiplication is a computational kernel of: (1) density matrix purification
\cite{hist-mcweeny, pur-pm,pur-n,pur-ntc,pur-h,pur-m, m-accPuri, dmu-jm}, (2)
density matrix minimization \cite{dmu-jm, dmm-lnv, dmm-ms,dmm-c, dmm-lojh,
dmm-sshw, dmm-shtjmojrpthc} (3) density matrix perturbation theory
\cite{purper-nw, per-nc, per-nwc}, (4) computation of interior eigenpairs of
potential matrices \cite{interior-wz, interior-xyhz, interior-rubenssonZahedi},
and (5) time--dependent response calculations \cite{response-chjtojrphs,
izmaylov:224105, kussmann:204103, lucero}.  Since matrices occurring in
electronic structure calculations often have a natural blocked structure arising
from local atom centered basis functions, performance can be dramatically
improved by using a blocked data structure \cite{m-rrs, CPC_128_93,
matrix-ssbrh, matrix-bmg}. In addition, the sparse blocked matrix--matrix
multiply is important in many other fields as for example the evaluation of
matrix functions including the matrix exponential, the matrix inverse
\cite{inch-bkt}, inverse factorizations \cite{hist-lowdin, inch-n, inch-jhjoh,
rubenssonBockHolmstromNiklasson}, and multigrid methods \cite{Banks_Sparse}
where local blocking may occur.

CPUs and memory of modern computer architectures work at different speeds. Main
memory operates at lower speeds to reduce the price and power consumption of
computers. Direct access to main memory causes the CPU to stall for several,
sometimes hundreds of cycles. To achieve decent computer performance cache
memory that store frequently accessed data was introduced into modern CPU
designs. Cache memory works at speeds comparable with the CPU but typically has
only a size of about 0.1\% to 1\% of computer main memory. The cache stores data
in chunks, so--called cache lines, which are usually on the order of tens of
bytes long.  A memory manager controls storing and evicting data from these
cache lines using heuristics (a typical algorithm called Least Recently Used
(LRU), evicts the least recently used cache line when trying to store a new one)
and it prefetches subsequent lines when sequential memory access is detected or
explicit machine language instructions are given to the CPU.  When implementing
a data intensive algorithm, both of these cache features -- LRU eviction and
hardware prefetching -- can be used to optimize performance.

Previous research into ordering effects for the matrix--matrix multiply has been
focused on either the dense or the very sparse case.  In the dense case, block
recursive algorithms have drawn much attention \cite{LNCS_3911_1042,
LAA_417_301, 305231, Gustavson_recursion}. In the case of sparse matrices,
\citet{Toledo:IMSPSMVM} found better than $2 \times$ speedups with the use of
locality enhancing orderings based on space filling curves.  Here,  we are
interested in the intermediate case of matrix--matrix multiplication involving
sparse matrices with local blocking, as occurs with local basis functions,
finite element methods \cite{strout2001rls} or reordering schemes
\cite{Li99ascalable, bai91direct, 762822}.

The performance of blocked matrix--matrix multiplication depends on (1) the
performance of block operations and (2) how blocks are stored and accessed.
Block operations can be delegated to some standard linear algebra library
optimized for the particular platform \cite{url:lapack, url:gotoblas, url:atlas,
url:mkl, url:acml}. Here, we focus on (2), exploring the effects of orderings
for small blocks, consistent with our interest in the intermediate case of
locally blocked sparse matrices.

This article is organized as follows: In section \ref{section:data_locality} we
describe in more detail the locality issues we are addressing. In section
\ref{section:results} we discuss our results and finally, in section
\ref{section:conclusions} we conclude.

\section{Data locality}
\label{section:data_locality}

Data locality is known to be important for achieving good performance of matrix
multiplication on modern computer architectures \cite{Toledo:IMSPSMVM}.  In
addition, Translation Lookaside Buffer (TLB) misses can significantly impact
performance \cite{url:gotoblas}. In this study we will focus on the locality
problem.

\subsection{Effects of ordering on performance}
\label{section:ordering}

We divide an $N \times N$ matrix into submatrices of size $b \times b$, where
$b$ is the blocksize. We choose $b$ so that the resulting blocked matrix will
consist of $n \times n$ submatrices. We assume for simplicity that $n \, b = N$.
The matrix product can be written as a combination of submatrix products,

\begin{equation}
  \label{eq:matrix_product}
  C_{ij} = \sum_{k = 1}^{n} A_{ik} B_{kj} \mbox{\hspace{1cm}} \left\{ i, j \in \left[ 1..n \right] \right\}.
\end{equation}

\noindent
The $n^{3}$ products may be evaluated in any order and we are free to choose the
precise ordering of execution and the allocation of the matrix blocks. Algorithm
\ref{alg:multiply} illustrates this point.

\begin{algorithm}
  \caption{General matrix multiplication algorithm}
  \label{alg:multiply}
  \begin{algorithmic}
    \STATE Allocate $n^{2}$ submatrix blocks for $A$, $B$, and $C$.
    \FORALL{$(i, j, k)$ in blocked matrix product of eq.~(\ref{eq:matrix_product})}
      \STATE Multiply blocks $A_{ik}$ and $B_{kj}$.
      \STATE Add result to block $C_{ij}$.
    \ENDFOR
  \end{algorithmic}
\end{algorithm}

\begin{figure}
  \includegraphics[width=0.9\linewidth]{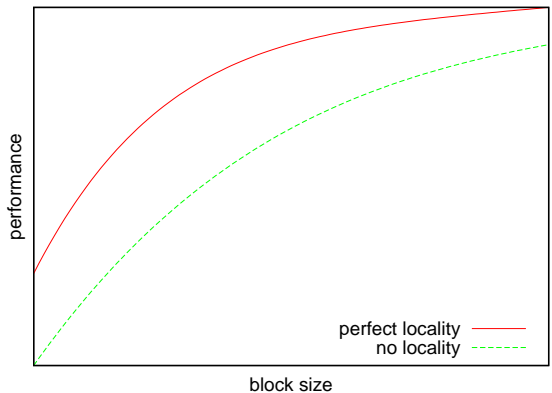}
  \caption{A qualitative illustration of the performance of a blocked matrix
  multiplication with perfect locality and no locality.}
  \label{fig:qualitative}
\end{figure}

Considering that we use a highly optimized matrix multiplication function on the
block level, what performance gain, if any, can we hope to achieve by arranging
the blocks in a particular order? Clearly, we can construct the two limiting
cases easily, ``perfect locality'' and perfect non-locality or ``no locality''.
We construct two tests: In the first test we multiply the same 2 blocks (e.g.
$A_{11}$ and $B_{11}$ to get $C_{11}$) $n^{3}$ times; in this way perfect
locality is obtained since the computer's memory manager can keep the three
submatrices in cache throughout the whole operation. In the second test we
randomize the multiplication and the allocation order of the blocks. This will
break most of the locality since memory prefetches are only possible in the rare
event that two blocks are close together in memory and the time between two
multiplication steps in which a particular block is reused is very long and will
almost certainly lead to a cache miss. If ordering effects are significant for
performance, we expect to find results qualitatively similar to those shown in
Fig.~\ref{fig:qualitative}.

\subsection{Space filling curves for optimized locality}

It is well known that ordering matrix elements in memory along locality
preserving space filling curves improves the performance of matrix operations
due to memory subsystem design issues \cite{Jin:USFCCR}. \citet{LAA_417_301}
applied this idea to dense matrix--matrix multiplication. They devised a block
recursive scheme which allocates the matrix elements along a Peano curve
\cite{Peano:MA_36_157} and reorders the multiplications of matrix
\emph{elements} to optimize locality.  They point out that such a scheme is
cache oblivious and platform independent.  Compared to the standard library MKL
\cite{url:mkl}, Bader and Zenger find that the reordered matrix--matrix
multiplication yields competitive performance when processor specific
optimization techniques, e.g. Intel's Streaming Single Instruction, Multiple
Data Extensions (SSE), are turned off.  In the following we will apply the idea
of improving locality by ordering to blocked matrices with blocks larger than
single matrix elements.

\begin{figure}
  \includegraphics[width=0.9\linewidth]{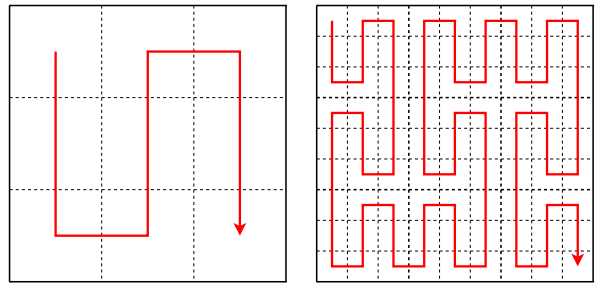}
  \caption{Left panel: The Peano curve ordering of the elements of a $3 \times
  3$ matrix block. Right panel: Recursive construction of Peano curve ordering.}
  \label{fig:peano_block}
\end{figure}

Figure~\ref{fig:peano_block} illustrates the recursive construction of the Peano
curve ordering of the matrix elements. This defines the matrix element index
ordering and also the ordering of the matrix elements in memory. The Peano
ordering does not uniquely define a multiplication order and we chose the
multiplication order that maximizes locality according to Bader and Zenger
\cite{LAA_417_301, LNCS_3911_1042}.

\subsection{Temporal vs. spatial locality}
\label{section:locality}

Locality can be divided into two types: spatial and temporal locality. Spatial
locality is the kind of locality one achieves by allocating data contiguously in
memory. Such locality takes advantage of the hardware prefetch -- after
operations on a block are finished, the next one can be found ready in the
cache. Temporal locality on the other hand means that data needed is already
present in cache because it was used in a previous computational step and does
not have to be loaded from memory. The impact of this locality is related to the
cache line management algorithm (e.g. LRU). From a programmer's point of view,
spatial locality concerns may influence the choice of the submatrix allocation
method, whereas temporal locality concerns may influence the choice of execution
order of the matrix multiplication.

It is reasonable to assume that by ordering the blocked matrix product along a
Peano curve, we optimize both spatial and temporal locality, just as is the case
for single matrix elements \cite{LNCS_3911_1042, LAA_417_301}. In the following
we refer to the case in which both spatial and temporal locality are optimized
along a Peano curve as ``temporal and spatial locality''. We want to separate
the two types of locality and understand how each of them affects performance.

We can destroy spatial locality by avoiding any kind of contiguous memory
allocation during the matrix block allocation. Elements within the matrix blocks
are of course still allocated in contiguous memory since we want to be able to
multiply matrix blocks by calling a standard generalized matrix--matrix multiply
({\tt gemm}). However, we randomize the allocating order of the submatrix
blocks, which makes it unlikely that two consecutive blocks are close to each
other in memory, and break in this way contiguous allocation on the inter--block
level. Since allocation order does not affect the multiplication order, and
therefore temporal locality, we can measure the effect of temporal locality by
itself and compare with our result for full Peano curve ordering. In the
following we will refer to this non--contiguous case as ``temporal locality''.
We expect the performance to lie somewhere between the 2 idealized curves
indicated in Fig.~\ref{fig:qualitative}.

\section{Results}
\label{section:results}

\begin{figure}
  \includegraphics[width=0.9\linewidth]{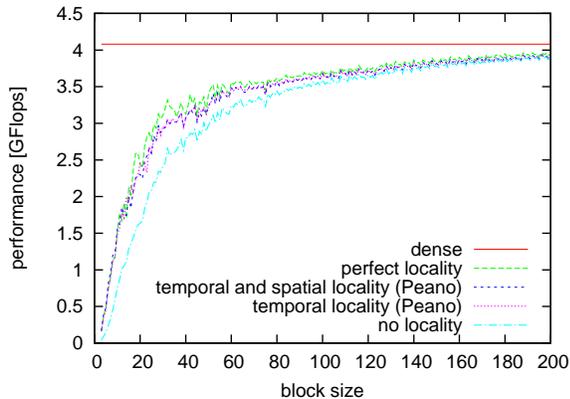}
  \caption{Opteron 248: Comparison of the blocked matrix multiplication
  performance for different ordering. The dense case is shown as reference and
  represents the performance of a dense matrix multiplication.}
  \label{fig:fod_performance}
\end{figure}

\begin{figure}
  \includegraphics[width=0.9\linewidth]{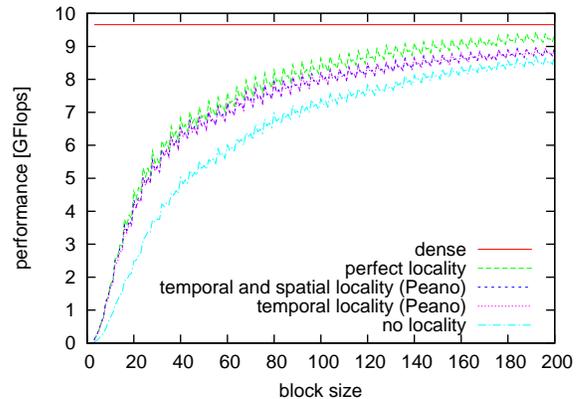}
  \caption{Xeon Woodcrest: Comparison of the performance of Peano curve ordering
  and Peano curve multiplication ordering with no spatial locality.}
  \label{fig:dk_performance}
\end{figure}

\begin{figure}
  \includegraphics[width=0.9\linewidth]{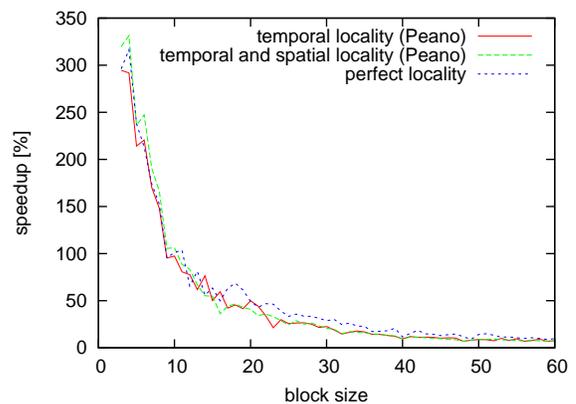}
  \caption{Opteron 248: Speedup of blocked matrix multiplication compared to the
  case of no locality. Speedup is given in percentage and a speedup of 100\%
  means a doubling of performance.}
  \label{fig:fod_speedup}
\end{figure}

\begin{figure}
  \includegraphics[width=0.9\linewidth]{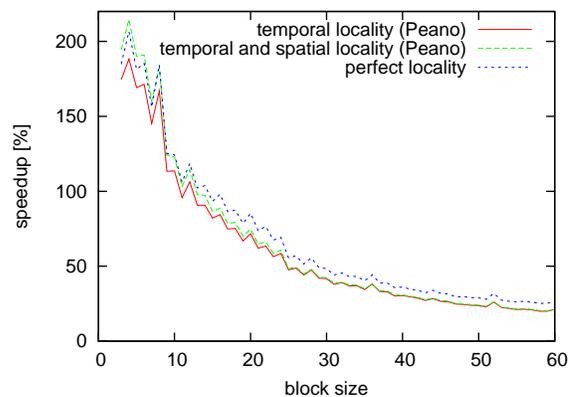}
  \caption{Xeon Woodcrest: Speedup of blocked matrix multiplication.}
  \label{fig:dk_speedup}
\end{figure}

In the previous section we discussed four different strategies of computing a
blocked matrix--matrix product with different data locality features: ``perfect
locality'', ``temporal and spatial locality'', ``temporal locality'', and ``no
locality''. Here we present the performance for the blocked matrix--matrix
multiplication with these locality features for block sizes ranging from 3 to
200, on two different computer architectures. All calculation were performed on
a single CPU.

The results shown in Fig.~\ref{fig:fod_performance} were obtained on an AMD
Opteron 248 system clocked at 2.2 GHz with 8 GiB\footnote{We denote memory sizes
in units if KiB, MiB, and GiB, which represent $1024^{1}$, $1024^{2}$,
$1024^{3}$ bytes respectively.} of main memory, using the GotoBLAS
\cite{url:gotoblas} library. The results shown in Fig.~\ref{fig:dk_performance}
were obtained on an Intel Xeon Woodcrest 5150 system clocked at 2.66 GHz with 8
GiB of main memory, using GotoBLAS \cite{url:gotoblas}.

We generate 2 random square matrices of size $N \times N$, where $2000 \leq N
\leq 6000$. These matrices are blocked into submatrices of size $b \times b$,
the block size.  The type of test determines the allocation method of the
submatrix blocks and the execution order of the multiplication. We time the
multiplication with the CPU time as reported by the operating system. The
performance is defined as $P = \left( 2N^{3}+4N^{2} \right) / t$, where $t$ is
the CPU time measured, taking into account one multiplication and one addition
per term in eq.~(\ref{eq:matrix_product}), one memory read operation per matrix
element, and one write operation per matrix element of the $C$ matrix. We repeat
each test 30 times to average out any fluctuations in the timing.

On both platforms, we find that ordering has a profound effect on performance.
The difference between a blocked multiplication with perfect locality and
without locality is significant (see section \ref{section:ordering} for a
definition of the terms ``perfect locality'' and ``no locality''). This point is
emphasized in Figs.~\ref{fig:fod_speedup} and \ref{fig:dk_speedup}, which show
the relative speedup achieved over the case of ``no locality''.  The performance
of the dense matrix test is independent of the matrix size for $N > 2000$. As a
reference we also show in Figs.~\ref{fig:fod_performance} and
\ref{fig:dk_performance} the performance of the {\tt gemm}() library call in the
large matrix limit. The ``dense'' result has to be seen as an upper limit of
what can be achieved for large matrices. Due to the fact that libraries such as
GotoBLAS tune their performance with regards to large matrices, we should expect
a drop in performance for small matrices. This is the reason why our results
indicate that a blocked approach becomes less efficient as the block size
decreases.

We clearly find that full Peano curve ordering achieves a performance which is
close to perfect locality.  This agrees with earlier findings by
\citet{LAA_417_301}. In the case of temporal locality, we find that the matrix
multiplication achieves near perfect locality performance as well (``temporal
locality'' refers to the case where we separately allocate submatrix blocks so
that they are not contiguous in memory, see section~\ref{section:locality}).
The relative speedup achieved by temporal ordering is almost identical to the
speedups achieved by Peano curve ordering or the case of perfect locality. This
is in contrast to the results for single matrix elements. This shows that
spatial locality becomes quickly irrelevant as the blocksize increases. We
believe that this result has not been appreciated until now.

\section{Discussion}

A closer look at how modern computer memory managers operate reveals that one
might expect our finding that performance of matrix--matrix multiplication is
controlled by temporal locality, and not spatial locality. As pointed out in the
introduction, the cache is organized in cache lines of fairly limited size. In
the case of the Opteron 248, a cache line contains 64 bytes.  The Opteron's
memory manager will prefetch cache line $n+3$ when it notices access to cache
line $n$, followed by access to cache line $n+1$, which is true also for the
case of a descending access pattern.  The total size of a typical cache is much
larger than the size of a single cache line and is typically on the order of
0.1\% to about 1\% of main memory. In the case of the Opteron 248, the size of
cache is 1 MiB. The size of cache of the Xeon Woodcrest architecture on the
other hand is 4 MiB, which is, at least partly, responsible for the delayed rise
of the multiplication performance when compared to the Opteron. To remind the
reader, we refer to spatial locality as having to do with prefetching and
temporal locality as referring to data being reused and therefore already
present in cache. We conclude that cache line size and memory prefetch are the
relevant features regarding spatial locality and total cache size is what
matters for temporal locality.  Given that a cache line is very small we do not
expect hardware prefetch effects and therefore spatial locality to matter for
the submatrix sizes we studied.  Temporal locality, on the other hand, is
important up to relatively large submatrix blocks given the different length
scale of the total cache size.

\section{Conclusions}
\label{section:conclusions}

We find that ordering effects for blocked matrix multiplications is important
for performance confirming earlier findings by other researchers
\cite{LNCS_3911_1042, LAA_417_301, CPC_128_93}. The performance gain is due to
increased locality. By breaking down the block locality into two types, spatial
and temporal locality, we find that temporal locality gives near perfect
locality by itself and that spatial locality can be neglected. This has
important implications for the implementation of any blocked matrix
multiplication method. The programmer should worry about the execution order of
the multiplication, but can safely ignore any concerns regarding contiguous
memory allocation. This allows for greater flexibility for blocked matrix data
structures.

\section{Acknowledgments}

The computational work for this study was performed on the ``Field of Dreams''
cluster of the Theoretical Division at Los Alamos National Laboratory and at the
``Horseshow'' cluster of the Danish Center for Scientific Computing (DCSC),
University of Southern Denmark. We gratefully acknowledge the support of the US
Department of Energy through the LANL LDRD/ER program for this work. We would
like to thank Denis Dimick, George Brehm, and Travis Peery for many insightful
discussions. Finally, we would like to thank the Ten Bar Caf\'{e} for its
impeccable service and stimulating atmosphere.

\bibliography{peano}
\bibliographystyle{apsrev}
%\bibliographystyle{siam}
%\newpage %achemso
%\listoffigures\newpage

\end{document}